\documentclass[aps,prb,twocolumn,superscriptaddress]{revtex4-2}

\usepackage[bookmarks=false,colorlinks=true,urlcolor=blue,citecolor=blue,linkcolor=blue]{hyperref}

\usepackage{cancel}
\usepackage{xspace}
\usepackage{tikz}
\usetikzlibrary{calc,arrows}

\usepackage{amsmath,amssymb,bm}
\usepackage{gensymb}
\usepackage[normalem]{ulem}
\usepackage{booktabs}

\usepackage{graphicx}

\newcommand{\mycomment}[1]{}

\newcommand{\markup}[1]{{#1}}

\newcommand{\be}{\begin{equation}}
\newcommand{\ee}{\end{equation}}
\newcommand{\bea}{\begin{eqnarray}}
\newcommand{\eea}{\end{eqnarray}}
\newcommand{\beal}{\begin{align}}
\newcommand{\eeal}{\end{align}}
\newcommand{\bes}{\begin{equation} \begin{split}}
\newcommand{\ees}{\end{split} \end{equation}}

\newcommand{\f}{\frac}

\newcommand{\qv}{\vec{q}}

\newcommand{\Rv}{\vec{R}}

\newcommand{\uv}{\vec{u}}
\newcommand{\vv}{\vec{v}}

\newcommand{\Sv}{\vec{S}}

\newcommand{\av}{\vec{a}}

\newcommand{\THUMP}{T_{\rm hump}\xspace}

\newcommand{\RN}[1]{%
\textup{\uppercase\expandafter{\romannumeral#1}}%
}

\usepackage{contour}

\definecolor{taylorswift}{rgb}{0.0862745098,0.4666666667,0.3411764706}
\definecolor{fearless}{rgb}{0.8862745098,0.6117647059,0.2823529412}
\definecolor{speaknow}{rgb}{0.4588235294,0.2274509804,0.4980392157}
\definecolor{red}{rgb}{0.6509803922,0.1254901961,0.2705882353}
\definecolor{TS1989}{rgb}{0.1803921569,0.6,0.9764705882}
\definecolor{reputation}{rgb}{0.1450980392,0.1490196078,0.1529411765}
\definecolor{lover}{rgb}{0.8392156863,0.2117647059,0.5529411765}

\tikzset{>=latex}

\begin{document}

\title{The Finite-Temperature Behavior of a Triangular Heisenberg Antiferromagnet}

\author{Cecilie Glittum}
\affiliation{Helmholtz-Zentrum Berlin für Materialien und Energie GmbH, Hahn-Meitner-Platz 1 14109 Berlin, Germany}
\affiliation{Dahlem Center for Complex Quantum Systems and Fachbereich Physik, Freie Universität Berlin, 14195 Berlin, Germany}
\author{Olav F. Sylju{\aa}sen}
\affiliation{Department of Physics, University of Oslo, P.~O.~Box 1048 Blindern, N-0316 Oslo, Norway}

\begin{abstract}
We investigate the classical antiferromagnetic Heisenberg model on the triangular lattice with up to third-nearest neighbor exchange couplings using the Nematic Bond Theory. This approach allows us to compute the free energy and the neutron scattering static structure factor at finite temperatures. We map out the phase diagram with a particular emphasis on finite-temperature phase transitions that break lattice-rotational symmetries, spiral spin liquids and the broad specific heat hump that is ubiquitous in the antiferromagnetic 120{\degree} phase. We identify this specific heat hump as signaling the onset of an exponentially increasing correlation length. Further, we map out the temperature of the specific heat hump and the transition temperatures of the symmetry-breaking transitions throughout the exchange-coupling space. Along the line $J_3 = J_2/2$, the Fourier-transformed exchange coupling exhibits a degenerate ring-like minimum, giving rise to spiral spin liquid behavior at intermediate temperatures. We investigate the structure factor of the spiral spin liquid as function of $J_2$ and identify the corresponding low-temperature order, which coincides with the single-$\qv$ spiral states of maximum spin-wave entropy along the degenerate ring.
\end{abstract}

\maketitle

\section{Introduction}

\begin{figure}
\centering
\includegraphics[width=0.5\textwidth]{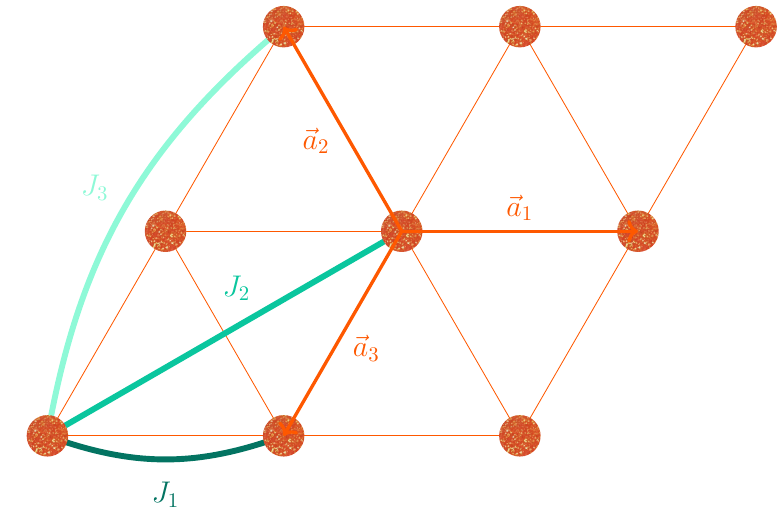}
\caption{Triangular lattice with first- ($J_1$), second- ($J_2$) and third-nearest ($J_3$) neighbor interactions.  $\av_1=(1,0)$, $\av_2 = (-1/2, \sqrt{3}/2)$ and $\av_3 = (-1/2, -\sqrt{3}/2)$ are the triangular lattice vectors.}
\label{fig:lattice}
\end{figure}

The antiferromagnetic (AF) Heisenberg model on the triangular lattice is the eldest daughter of frustrated magnetism.
Beyond its fundamental theoretical importance, it has received renewed interest in recent years, driven by the discovery of new materials. In particular, the $S=1/2$ delafossites NaYbSe$_2$, KYbSe$_2$ and CsYbSe$_2$ are layered triangular lattice materials with negligible interlayer couplings. They are believed to be described by a $J_1$-$J_2$ AF \textit{quantum} Heisenberg model (see Fig.~\ref{fig:lattice} for definitions of exchange couplings) with $J_2$ in proximity to a suggested quantum spin liquid (QSL) region~\cite{Scheie2024, Scheie2024_2,Xie2023}.
Recent studies of $S=2$ h-Lu$_{0.3}$Y$_{0.7}$MnO$_3$ and $S=5/2$ h-Lu$_{0.47}$Sc$_{0.53}$FeO$_3$ have further highlighted the interplay of frustration, quantum effects, and potential applications of triangular Heisenberg magnets in spintronics and quantum information~\cite{Yano2024}.

Although these systems are inherently quantum mechanical, the classical Heisenberg model emerges as the $S\to \infty$ limit of the quantum spin model. This limit is not only appropriate for large-spin materials, but serves also as a guide for locating regions where different classical orders are nearly degenerate and the ordering temperature is suppressed.
In such regions, it is natural to expect quantum fluctuations to play an important role in the corresponding quantum model at low temperatures. They may destabilize classical order and potentially induce quantum‑disordered phases, including QSLs or other exotic states.
For example, the QSL regime of the quantum $J_1$-$J_2$ AF Heisenberg model lies close to a phase boundary in the classical model.

The classical Heisenberg model on the triangular lattice with up to third-neighbor interactions has a rich ground state phase diagram~\cite{Rastelli1980,Jolicoeur1990, Korshunov1993, Messio2011, Mohylna2022}.
Yet, the fate of these phases at finite temperatures remains largely unresolved. Although the Mermin-Wagner theorem precludes breaking of the global spin symmetry at finite temperatures~\cite{Mermin1966}, phase transitions can break discrete lattice symmetries, leading to, e.g., single-$\qv$ spiral states. To our knowledge, only a few exchange parameter points have been investigated as a function of temperature~\cite{Kawamura1984,Kawamura2010,Aoyama2020,Mohylna2022}.

Beyond the single-$\qv$ orders, the ferromagnetic (FM) counterpart of the $J_1$-$J_2$-$J_3$ model exhibits spiral spin liquid behavior~\cite{Glittum2021}. Similar behavior is also expected for the antiferromagnet close to the line $J_3 = J_2/2$, where there is a continuous ground state degeneracy of spiral states. Previously, two parameter points on this line were investigated, suggesting a spiral spin liquid being realized at intermediate temperatures with potential phase transitions into lattice-symmetry broken states at low temperatures~\cite{Mohylna2022}.

The classical triangular Heisenberg antiferromagnet thus remains a critical reference for understanding the effects of frustration, both as a standalone system and as a baseline for quantum models.
Using the Nematic Bond Theory (NBT)~\cite{Schecter2017, Glittum2021}, we map out the finite-temperature phase diagram of the model as function of second- and third-neigbor couplings, identifying specific heat behaviors, lattice-symmetry breaking phase transitions and spiral spin liquids.

We begin by introducing the model and method in Section~\ref{sec:model}. In Section~\ref{sec:results}, we present the results, with Section~\ref{sec:cv} focusing on the specific heat and critical temperatures in the full exchange coupling space. Section~\ref{sec:ring} investigates the finite-temperature behavior of the system in the spiral spin liquid regime.  Finally, we discuss our results and conclude in Sections~\ref{sec:discussion} and~\ref{sec:conclusion}, respectively.

\section{Model and Method \label{sec:model}}

\subsection{Hamiltonian}
We study the classical Heisenberg model on the triangular lattice with up to third-nearest neighbor couplings
\begin{equation}
H = J_1 \sum_{\langle i,j \rangle} \vec{S}_i \cdot \vec{S}_j
+ J_2 \hspace{-1.1mm} \sum_{\langle\langle i,j \rangle\rangle} \vec{S}_i \cdot \vec{S}_j
+ J_3 \hspace{-2.2mm} \sum_{\langle\langle\langle i,j \rangle\rangle\rangle} \vec{S}_i \cdot \vec{S}_j,
\end{equation}
where $\vec{S}_i$ are unit vectors. We will consider AF $J_1$ and use units where $J_1=1$ and $k_B=1$.
The corresponding exchange coupling in momentum space takes the form
\be
J_{\qv}  =   -3J_3 + (J_1 - 2J_3) A_{\qv} + 2J_3 A^2_{\qv} + (J_2 - 2J_3) B_{\qv},
\ee
where
\begin{equation}
\begin{split}
&A_{\qv} = \cos{q_1}+\cos{q_2}+\cos{q_3}, \nonumber \\
&B_{\qv} = \cos{(q_1-q_2)}+\cos{(q_2-q_3)}+\cos{(q_3-q_1)},\nonumber
\end{split}
\end{equation}
with $q_i \equiv \qv\cdot \av_i$ and triangular lattice vectors $\av_1=(1,0)$, $\av_2 = (-1/2, \sqrt{3}/2)$ and $\av_3 = (-1/2, -\sqrt{3}/2)$. We set the lattice spacing to unity.

\begin{figure}[]
\centering
\includegraphics[width=0.5\textwidth]{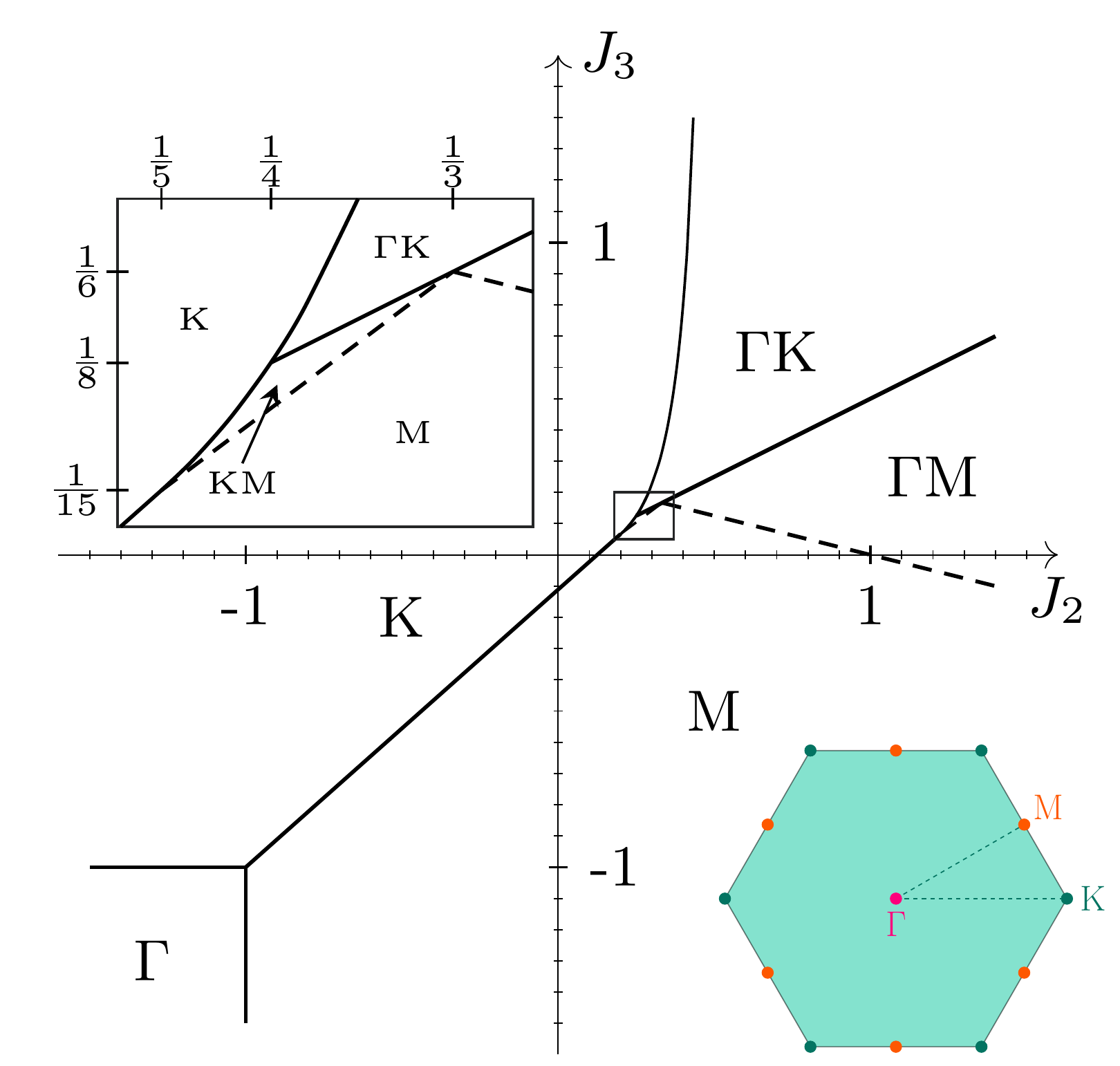}
\caption{Zero-temperature phase diagram with $J_1=1$~\cite{Rastelli1980}. Solid(dashed) lines indicate discontinuous(continuous) phase transitions. On the lower right, we show the first Brillouin zone of the triangular lattice with high-symmetry points. The thin dashed lines mark the lines $\Gamma$K and $\Gamma$M.}
\label{fig:phasediagram_zeroT}
\end{figure}

The characterization of the minima of $J_{\qv}$ was first made in Ref.~\onlinecite{Rastelli1980}, and their zero-temperature phase diagram is reproduced in Fig.~\ref{fig:phasediagram_zeroT}. The labelling of the phases denotes where in the Brillouin zone $J_{\qv}$ is minimal (see Fig.~\ref{fig:phasediagram_zeroT} for definitions of high-symmetry points).
For the pure nearest neighbor model, $J_{\qv}$ depends only on $A_{\qv}$. It is minimal on the K points in the Brillouin zone, and the ground state is the AF $120\degree$ state.
In contrast, $B_{\qv}$ is maximal at the K points and minimal at the M points. Thus for $J_3=0$, a sufficiently positive value of $J_2$ will cause the minimum to shift away from the K points to the M points.
The energy at M is lower than at K as long as
\be
J_2 > \f{1}{8} \left( 1+ 9J_3 \right)
\ee
or equivalently
\be
J_3 <  \f{1}{9} \left( -1 + 8 J_2 \right).
\ee
In particular, this means that for $J_3=0$($J_2=0$) the stripe phase (M), which breaks the lattice-rotational symmetry, becomes minimal for $J_2>1/8$($J_3 < -1/9$).
For both $J_2 <-1$ and $J_3<-1$ the minimum is at $\Gamma$, i.e. a FM state, and for large positive values of $J_2,J_3$ the minima lie on the lines $\Gamma$K or $\Gamma$M. For coupling values where K, M, $\Gamma$K and $\Gamma$M phases meet there is also a small region, see inset of Fig.~\ref{fig:phasediagram_zeroT},  where the minima lie on the KM lines on the Brillouin zone boundary.

The $\vec{q}$'s minimizing $J_{\qv}$ in the different regions are given by
\begin{gather}
\Gamma = (0,0), \; {\rm K} = (4\pi/3,0), \; {\rm M} = (0, 2\pi/\sqrt{3}),\\
\qv_{\Gamma {\rm M}} = (0, q^y_{\Gamma {\rm M}}), \; \qv_{\Gamma {\rm K}} = (q^x_{\Gamma {\rm K}}, 0), \; \qv_{\rm KM} = (q^x_{\rm KM},2\pi/\sqrt{3}),
\end{gather}
with
\begin{equation}
\begin{split}
q^y_{\Gamma {\rm M}} &= \frac{2}{\sqrt{3}}\arccos \left(\frac{-J_1 - J_2}{2J_2+4J_3}\right),\\
q^x_{\Gamma {\rm K}} &= 2\arccos \left(\frac{-3J_2+2J_3 + \sqrt{(3J_2+2J_3)^2-8J_1J_3}}{8J_3}\right),\\
q^x_{\rm KM} &= 2\arccos \left(\frac{3J_2-2J_3 + \sqrt{(3J_2+2J_3)^2-8J_1J_3}}{8J_3}\right),
\end{split}
\end{equation}
and their symmetry equivalents.
A corresponding ground state would be a single-$\qv$ spiral state
\begin{equation}
\Sv (\Rv) = \uv\cos{(\qv\cdot\Rv)} + \vv\sin{(\qv\cdot\Rv)},
\end{equation}
where $\uv$ and $\vv$ are orthonormal vectors defining the spiral plane. These states break the rotational symmetry of the lattice unless $\qv = \Gamma, {\rm K}$. For illustrations of the ordered spin configurations in each region of the phase diagram, see Ref.~\onlinecite{Rastelli1980}. In principle, multi-$\qv$ states might also be ground states, but only in special cases, as multi-$\qv$ states are generally difficult to normalize. For the model considered here, the M phase is known to also host a multi-$\qv$ ground state, the tetrahedral state, which is non-coplanar and preserves lattice symmetries~\cite{Messio2011}.

For $J_3=J_2/2$, the $B_{\qv}$-terms are cancelled, and $J_{\qv}$ is a quadratic polynomial in $A_{\qv}$ with the minimum forming a continuous ``ring'' for $J_2> 1/4$, on which the minimal $\qv$'s satisfy
\be
A_{\qv}  = \f{1}{2} \left( 1-\f{J_1}{J_2} \right).
\ee
This ring-minimum occurs where the $\Gamma$K phase borders the KM phase ($1/4 < J_2 \leq 1/3$) or the $\Gamma$M phase ($J_2 \geq 1/3$).

\subsection{Method}

To investigate finite-temperature properties of the model, we employ the NBT~\cite{Schecter2017}. This method provides a self-consistent treatment of spin correlations while enforcing the unit spin length on each lattice site. It allows us to compute the temperature dependence of the free energy~\cite{Glittum2021} and the static structure factor~\cite{Syljuasen2019} for large system sizes $L \times L$.
The NBT method focuses on $K^{-1}_{\qv}$ which is proportional to the scalar spin-spin correlation function and is written
\be
K^{-1}_{\qv} = \f{1}{J_{\qv} + \Sigma_{\qv} + \Delta},
\ee
where $\Sigma$ is the $\qv$-dependent self-energy and $\Delta$ is a control parameter. The self-energy is determined from a self-consistent equation that is derived from a diagrammatic expansion. For a detailed description of NBT, see Refs.~\onlinecite{Syljuasen2019,Glittum2021}.

The NBT diagrammatic expansions for the self-energy and the free energy are organized as large-$N_s$ expansions where $N_s=3$ is the number of spin components. The expansions are approximate, as vertex corrections and also some other diagrams of order $1/N_s$ are neglected.
Nevertheless, the method has previously proven to produce the correct phases and phase transitions~\cite{Glittum2023}, but it typically overestimates critical temperatures by 10-20\%~\cite{Syljuasen2019}.
The overestimation of critical temperatures is likely a consequence of the neglect of certain fluctuations in the $1/N_s$ expansion, meaning that all temperatures presented in this article should be understood as estimates.

Operationally, the NBT self-consistent equation is solved by iterations, typically starting from a random self-energy and a specific value of the control parameter $\Delta$. The equation is iterated until the self-energy converges. Then the temperature, free energy and structure factor are calculated.

\begin{figure}[]
\centering
\includegraphics[width=0.5\textwidth]{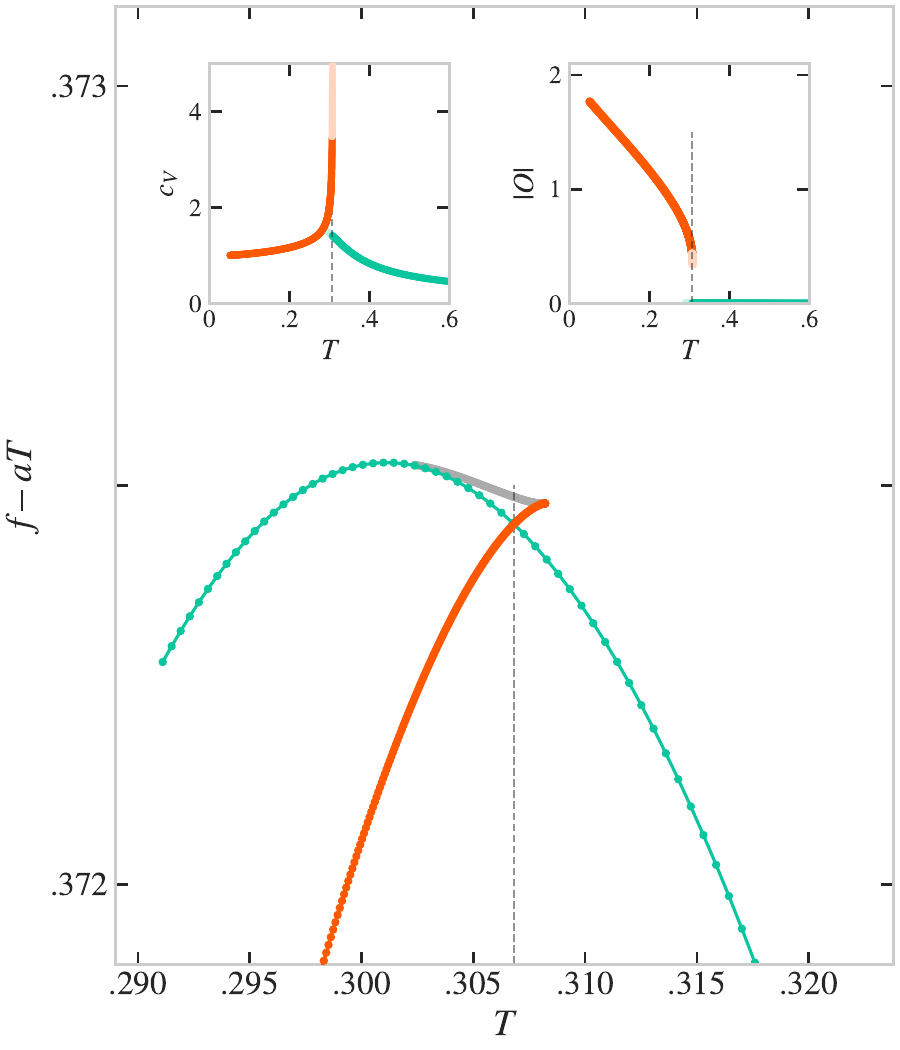}
\caption{Main panel: Free energy vs. temperature for $J_2=0.25$, $J_3=0$. For better visualization of the crossing of the free energy branches, a term $a T$ with $a=-1.64319$ has been subtracted from the free energy. Left inset: Specific heat vs. temperature computed separately for low- (orange) and high-temperature (green) branches. Right inset: The lattice-rotational symmetry breaking order parameter vs. temperature. In the insets, we have used fainter colors for points belonging to metastable states. The dashed vertical lines mark $T_c =0.30681$ where the free energy branches cross. $L=300$.}
\label{fig:freeexample}
\end{figure}

The free energy can be used to determine the thermodynamically stable states and phase transitions. An example is shown in Fig.~\ref{fig:freeexample} for parameter values $J_2=0.25$ and $J_3=0$ inside the M region (see Fig.~\ref{fig:phasediagram_zeroT}), where NBT produces a free energy having three branches.
The low-temperature branch (orange) is obtained by starting with a low value of $\Delta$ and a random self-energy. The self-consistent equations are then iterated until convergence, and the temperature and free energy are obtained.
Then $\Delta$ is increased slightly, but now the converged self-energy from the previous run is used as input instead of a random one. Generally the temperature increases as $\Delta$ increases.
This is repeated to produce the low-temperature branch. At a certain value of $\Delta$ the temperature begins to decrease even though $\Delta$ is increased, resulting in an intermediate branch (grey), until it ends abruptly, and the iterations converge to a point on the high-temperature branch (green) at another temperature. Increasing $\Delta$ further increases the temperature, giving more points on the high-temperature branch. One can also start by decreasing $\Delta$ from a high value. This produces again the high-temperature branch that continues downward in temperature until it ends abruptly and subsequent iterations converge to the intermediate or low-temperature branch.

\begin{figure}[]
\begin{center}
\includegraphics[width=0.5\textwidth]{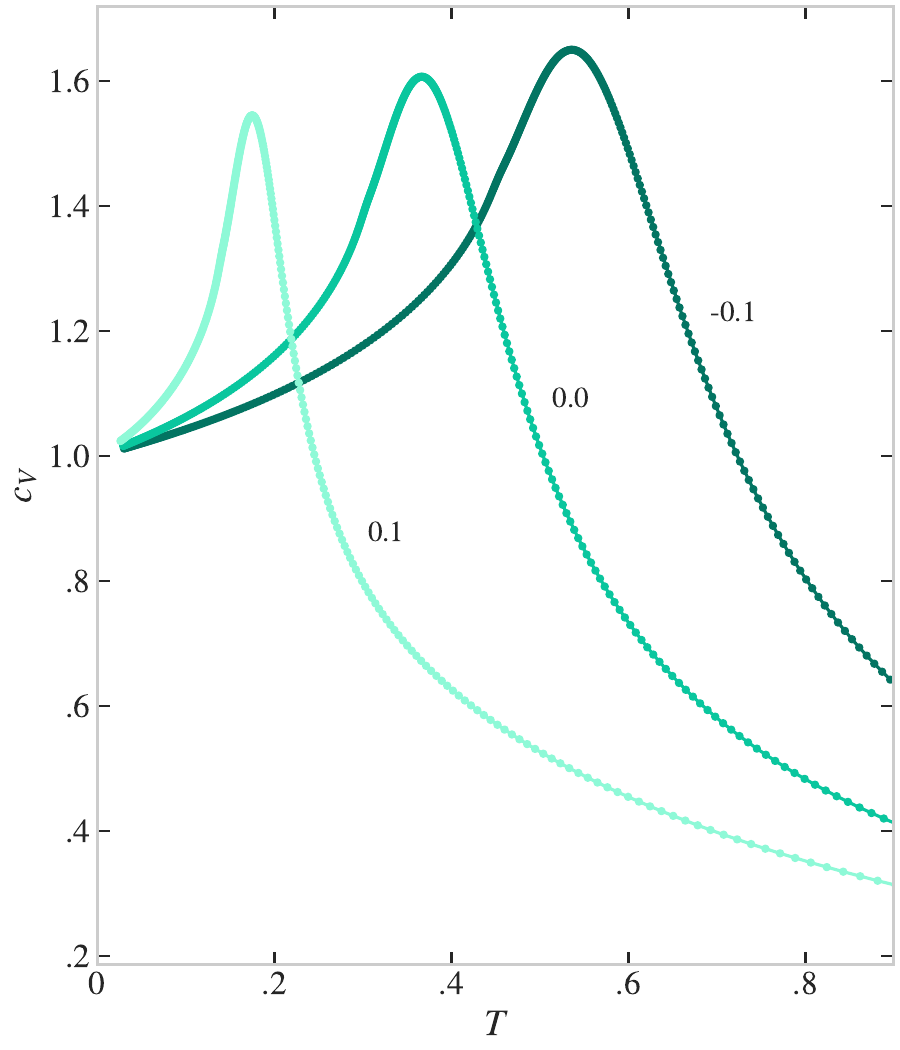}
\caption{Specific heat vs. temperature for different values of $J_2$ (indicated by the legends). $J_3=0$. $L=300$.}
\label{fig:humpJ2}
\end{center}
\end{figure}

\begin{figure}[]
\begin{center}
\includegraphics[width=0.5\textwidth]{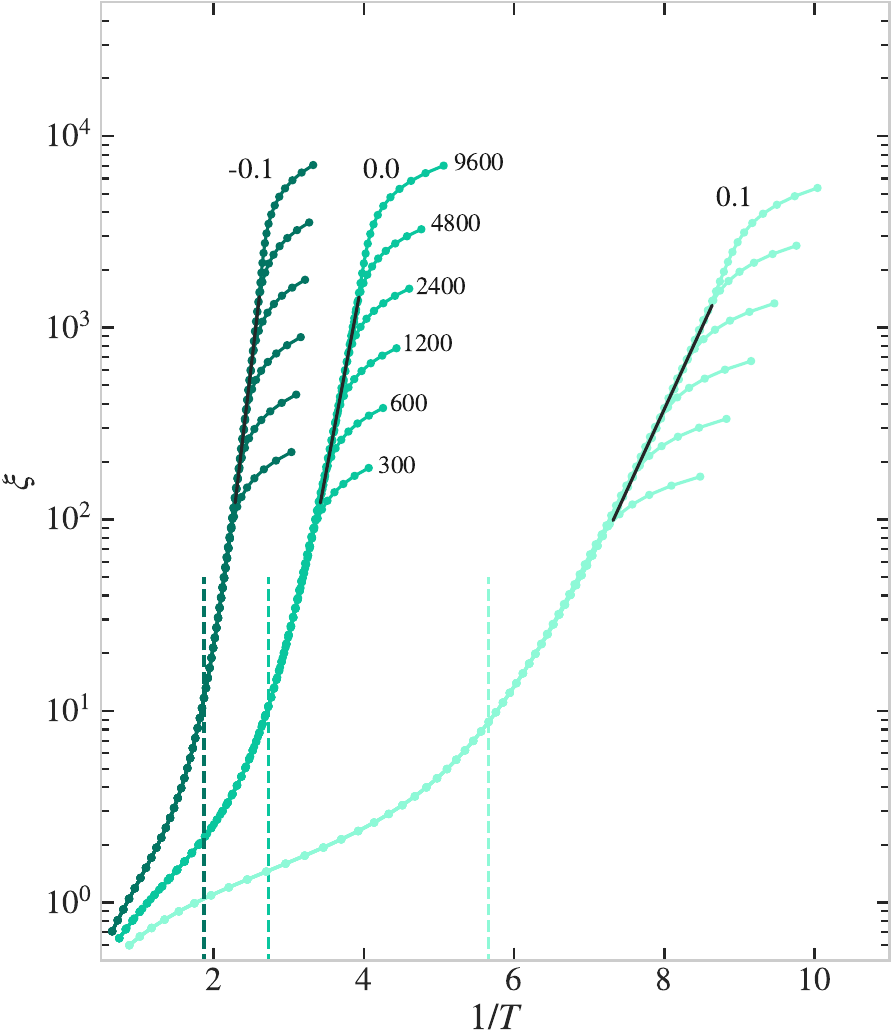}
\caption{Semilog plot of correlation length vs. inverse temperature (system size and $J_2$ indicated by the legends). $J_3=0$. The dashed vertical lines show $\THUMP$ for the biggest system size, $L=9600$. The straight lines have slopes $B=\{7.7,4.8,2.0\}$ for $J_2=\{-0.1,0,0.1\}$, respectively.
\label{fig:corrlength}}
\end{center}
\end{figure}

The thermodynamically stable state at a given temperature is the one with lowest free energy and negative curvature. The crossing point $T_c$ between the low- (orange) and high-temperature (green) branches marks therefore a phase transition. When the free energy branches approach the crossing point with different slopes, as is the case here, the phase transition is discontinuous having a latent heat. The parts of the free energy branches with negative curvatures that are not the lowest free energy for a given temperature indicate metastable states, and branches with positive curvature are unstable states.

From the free energy as a function of temperature, we can compute the specific heat separately for the free energy branches and glue them together at $T_c$, see left inset of Fig.~\ref{fig:freeexample}.
Note that for discontinuous phase transitions there is also an additional Dirac delta-function contribution to the specific heat at $T_c$ (not shown) which reflects the latent heat of the transition.

As the system considered in this paper is two-dimensional, breaking of the continuous  global spin rotational symmetry is precluded by the Mermin-Wagner theorem~\cite{Mermin1966}. NBT may still capture phase transitions associated with the breaking of \textit{discrete lattice symmetries} due to the $\qv$-dependence of the self-energy. Note that for a two-dimensional system, all single-$\qv$ phases should retain lattice inversion symmetry, as $\qv \to -\qv$ corresponds to a global spin rotation. On the other hand the $60\degree$ rotational symmetry of the triangular lattice can be broken, and is measured by the following order parameter
\begin{equation}
O = \frac{1}{L^2}\sum_{\qv}\langle \Sv_{\qv}\cdot\Sv_{-\qv}\rangle (e^{iq_1} + e^{i2\pi/3}e^{iq_2} + e^{i4\pi/3}e^{iq_3}),
\end{equation}
which is non-zero if the spin correlations differ along the distinct nearest-neighbor directions of the triangular lattice. We show this order parameter in the right inset of Fig.~\ref{fig:freeexample}, illustrating how the lattice-rotational symmetry is broken in states contributing to the low-temperature branch, but remains intact for high-temperature states.

\section{Results \label{sec:results}}

\subsection{Specific heat and phase transitions} \label{sec:cv}

For the pure nearest-neighbor antiferromagnet ($J_2 = J_3 = 0$), NBT gives only one free energy branch, and no phase transition at finite temperatures. The corresponding specific heat exhibits a broad asymmetric peak, which we will term a {\em hump}, at $\THUMP \approx 0.367$, see Fig.~\ref{fig:humpJ2} (middle curve). This hump becomes narrower and moves down in temperature when adding AF $0 < J_2 <1/8$, see Fig.~\ref{fig:humpJ2} (left curve). These specific heat curves all indicate $c_{V}(T=0)=1$. This is consistent with previous Monte Carlo simulations, and can be explained by the ordered ground state at $T=0$ having in total $2N-3$ energetically costly quadratic {\em amplitude} degrees of freedom (regardless of the linear dispersion). By the equipartition principle each quadratic amplitude degree of freedom contributes $\f{1}{2}$ to the heat capacity, giving a specific heat $c_V = \f{1}{N} \left( 2N-3 \right) \times \f{1}{2}  \to 1$ in the thermodynamic limit~\cite{Kawamura1984,Chalker1992}.

In Fig.~\ref{fig:corrlength}, we show the correlation length as a function of inverse temperature for the same three $J_2$ values as in Fig.~\ref{fig:humpJ2}. The correlation length is extracted by fitting the peak of the spin structure factor around the K point with a Lorenzian in the $\Gamma$K direction
\begin{equation}
S(\qv) \equiv \langle \Sv_{-\qv}\cdot \Sv_{\qv} \rangle = \frac{A}{(\qv-{\rm K})^2+\xi^{-2}}.
\end{equation}
The choice of direction is not important for low temperatures, as most weight of the peak lies close to the K point, where $J_{\qv}$ is circularly symmetric (to second order in the deviation $\qv - {\rm K}$).

It is seen from the curves in Fig.~\ref{fig:corrlength} that for temperatures below $\THUMP$ (dashed vertical lines), the correlation length increases exponentially
with inverse temperature, $e^{B/T}$, with $B$ a constant dependent on the exchange couplings. See caption of Fig.~\ref{fig:corrlength} for values of $B$.

\begin{figure}[]
\begin{center}
\includegraphics[width=0.5\textwidth]{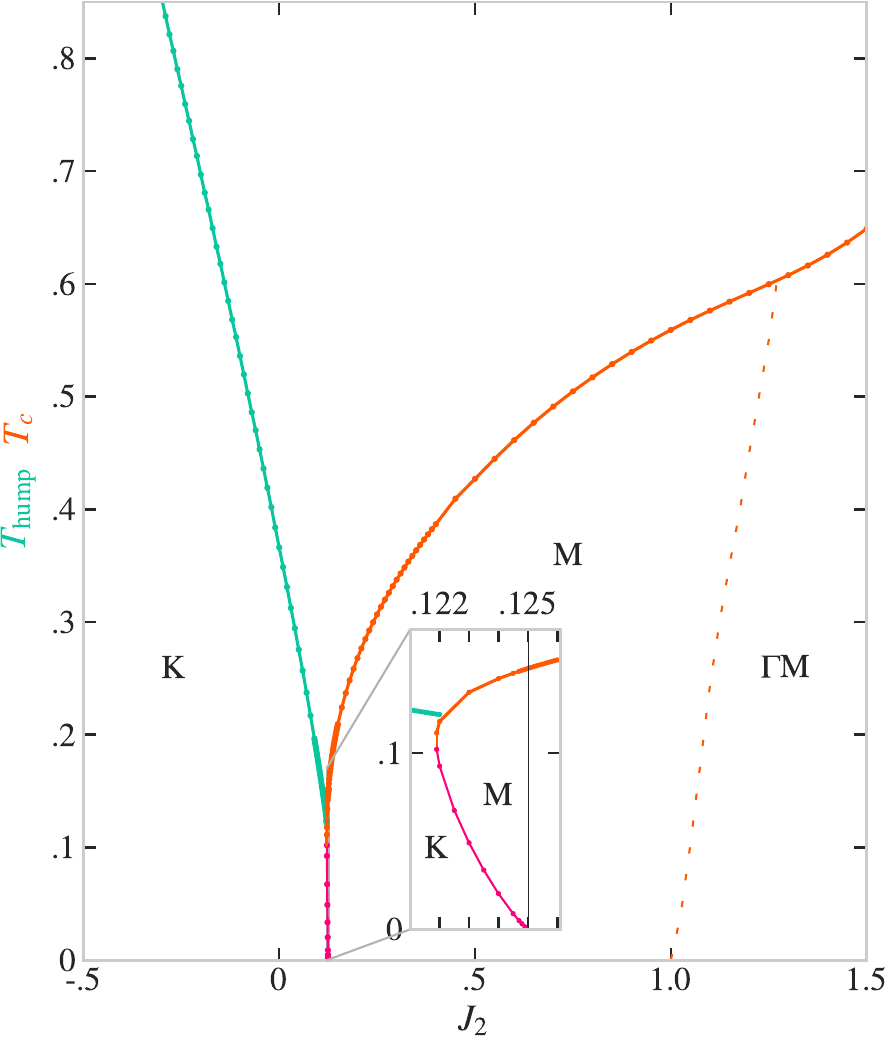}
\caption{$\THUMP$ (green) and $T_c$ (orange) vs. $J_2$ for $J_3=0$. The pink symbols show the lower $T_c$ associated with the phase transition between the ordered M and (disordered) K phases. Both orange and pink points represent discontinuous phase transitions. The orange dashed line indicates a crossover region where the ordering wave vector goes smoothly from M to $\Gamma$M as the temperature is lowered. $L=300$.}
\label{fig:J3zero}
\end{center}
\end{figure}

When $1/8 \leq J_2 \leq 1$, the minima of $J_{\qv}$ are on the M points, and the system shows a finite temperature discontinuous phase transition associated with the selection of one of the three inequivalent M points (single-$\qv$ stripe phase). Fig.~\ref{fig:freeexample} shows the free energy for a point in this region, $J_2 =0.25$, where the rotational symmetry-breaking order parameter is shown in the right inset.
Transition temperatures for other values of $J_2$ are given in Fig.~\ref{fig:J3zero}. Despite the minima of $J_{\qv}$ being on the K points for $J_2 < 1/8$, we find that the stripe phase is thermodynamically stable at intermediate temperatures also for the narrow region $J_2=\left[0.122,0.125 \right)$ (see inset of Fig.~\ref{fig:J3zero}). It extends down to very low temperatures at which it terminates in a discontinuous phase transition (pink symbols) into the disordered K phase. For comparison we have in Fig.~\ref{fig:J3zero} also plotted the temperature of the specific heat hump maximum. The two temperatures $\THUMP$ and $T_c$ both decrease and approach each other as $J_2$ gets close to $0.122$.

\begin{figure}[]
\begin{center}
\includegraphics[width=0.5\textwidth]{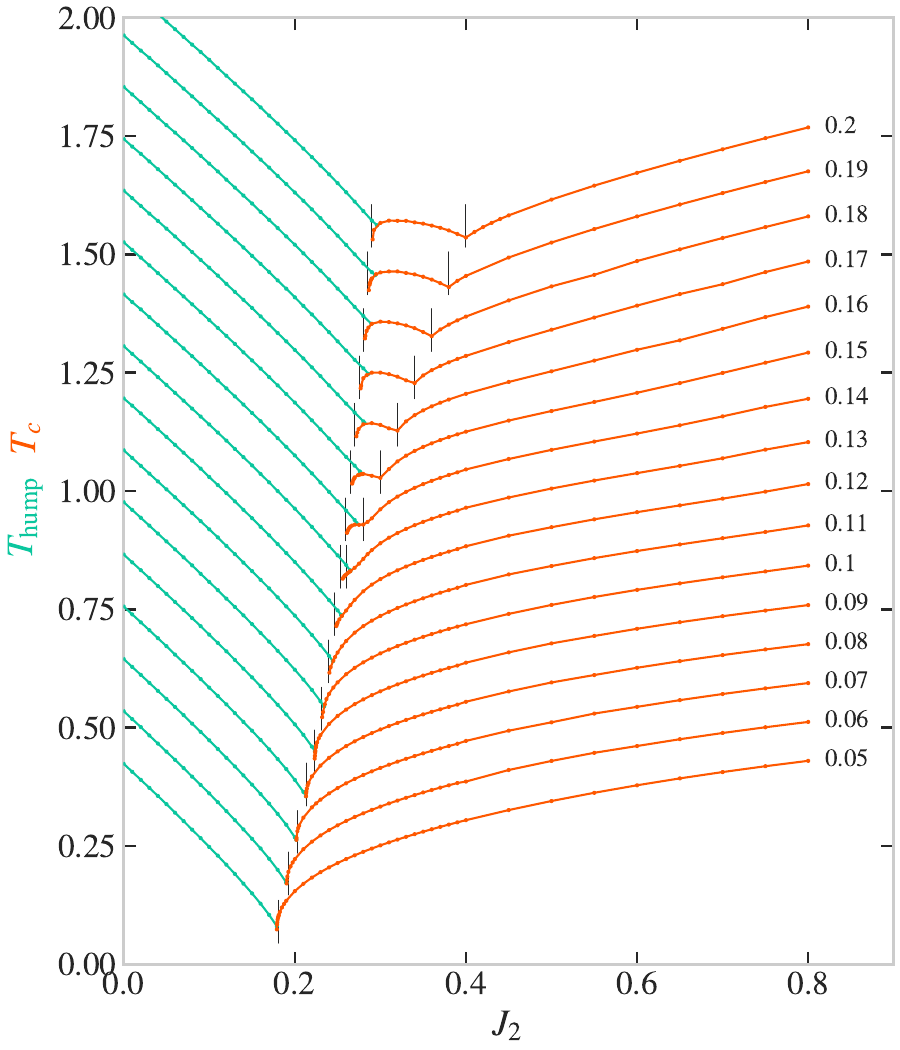}
\caption{$\THUMP$ (green) and $T_c$ (orange) vs. $J_2$ for different values of $J_3$ indicated by the numbers (right). Each curve is lifted by an amount $(J_3-0.05) \times 10$ to avoid overlaps. The vertical bars indicate the positions of the $T=0$ phase boundaries from Fig.~\ref{fig:phasediagram_zeroT}. All phase transitions are discontinuous. $L=300$.}
\label{fig:tcslices}
\end{center}
\end{figure}

\begin{figure}[]
\begin{center}
\includegraphics[width=0.5\textwidth]{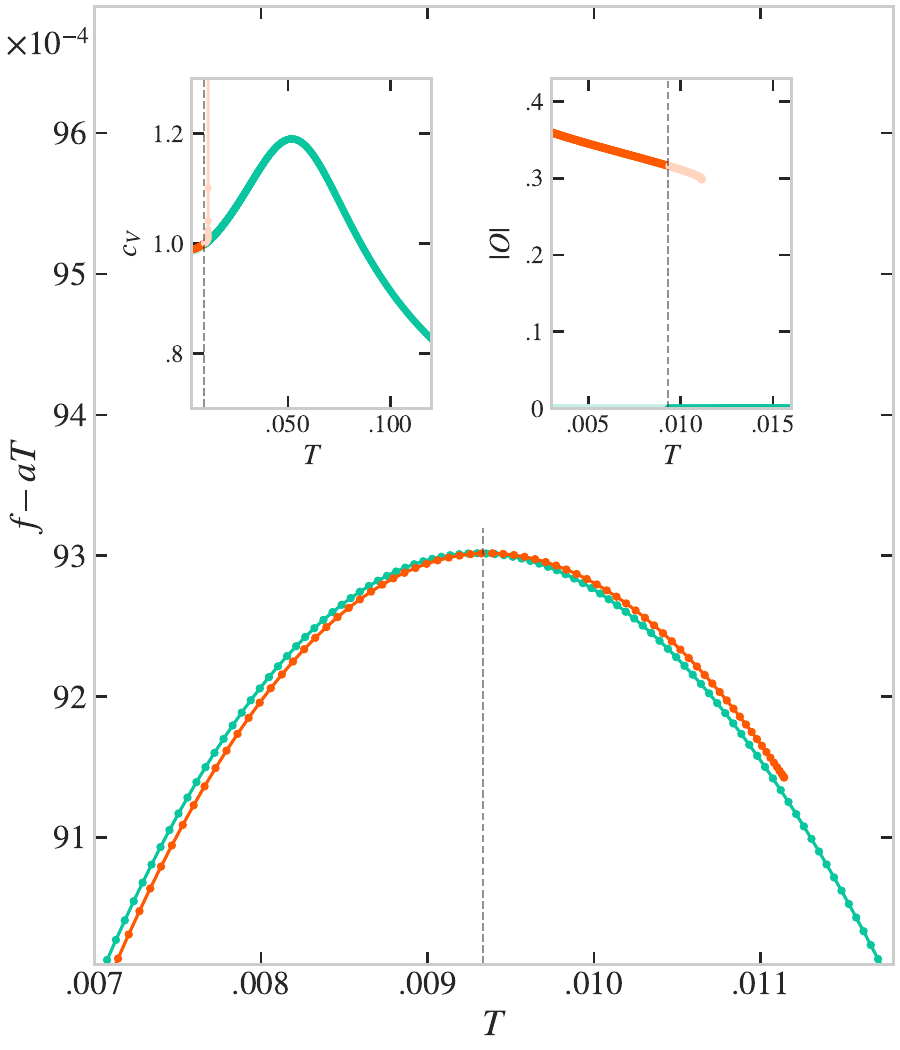}
\caption{Main panel: Free energy vs. temperature for $J_2=0.24$, $J_3=0.11$. For better visualization of the crossing of the two free energy branches, a term $ a T$ with $a=1.3355$ has been subtracted from the free energy. Left inset: Specific heat vs. temperature computed separately for the two branches over a wider $T$ region. Right inset: The lattice-rotational symmetry breaking order parameter vs. temperature for the two branches. In the insets, we have used fainter colors for points belonging to metastable states. The dashed vertical lines mark $T_c=0.009331$ where the two free energy branches cross. $L=300$.}
\label{fig:free}
\end{center}
\end{figure}

\begin{figure}[]
\begin{center}
\includegraphics[width=0.5\textwidth]{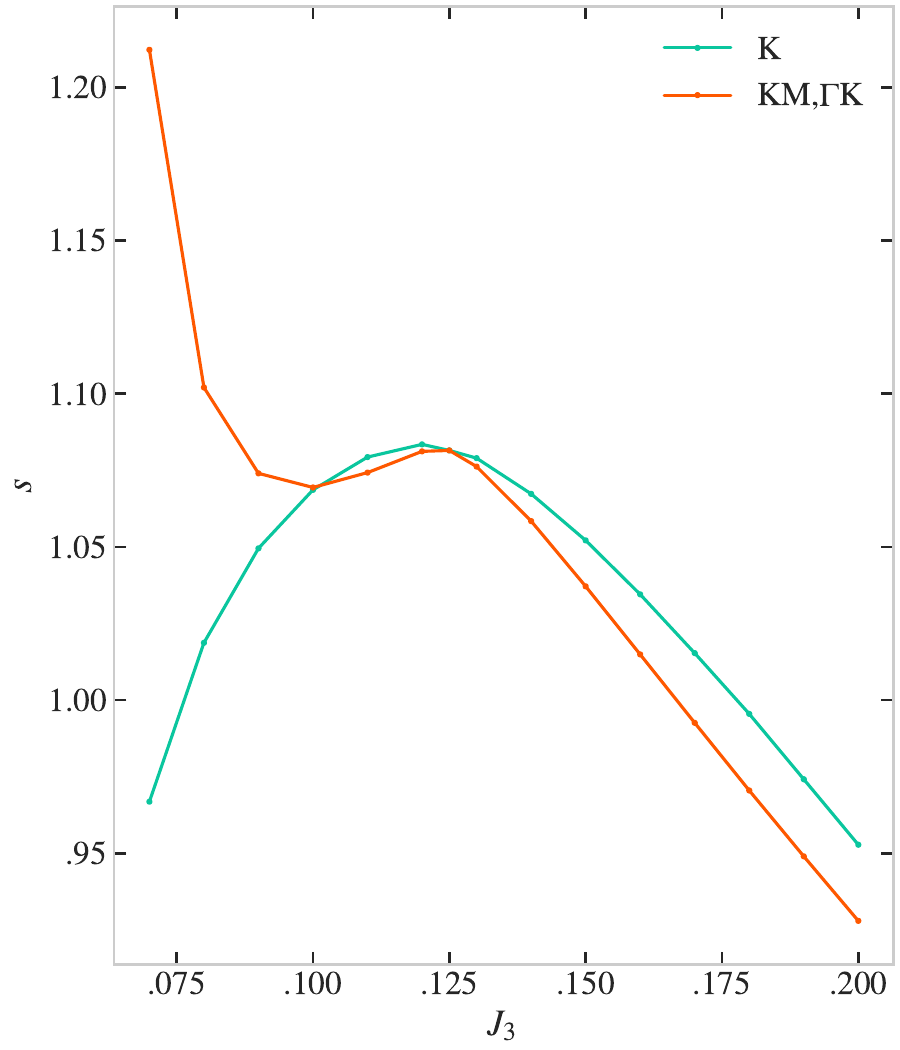}
\caption{$T=0$ spin-wave entropy for the K state (green) and KM or $\Gamma$K state (orange) along the K-KM and K-$\Gamma$K borders (see Fig.~\ref{fig:phasediagram_zeroT}). The points correspond to the $J_3$ values used in Fig.~\ref{fig:tcslices}. We have also included $J_3 = 1/8$ where all three states, and thus also their entropies, are equal. $L=300$.}
\label{fig:hump_entro}
\end{center}
\end{figure}

To see the specific heat hump temperatures and critical temperatures for finite positive values of $J_3$, we show in Fig.~\ref{fig:tcslices} $\THUMP$ (green) and $T_c$ for the highest temperature phase transition (orange) as functions of $J_2$ for a range of $J_3$ values. With the resolution used, one can clearly see cusp signatures in $T_c$, reflecting the $T=0$ phase boundaries from Fig.~\ref{fig:phasediagram_zeroT}.
For $J_3 \gtrsim 0.11$ there is a small $J_2$ region outside the K phase boundary (left vertical lines) for which the specific heat exhibits both a hump and a phase transition. For more details on this, see Fig.~\ref{fig:free} which shows the free energy, specific heat and order parameter at the specific point $J_2=0.24$, $J_3=0.11$. The free energy has two branches. The specific heat hump (left inset) is seen to come from the high-temperature branch (green) which reflects properties of the symmetric phase with a structure factor maximum at K, and occurs at a higher temperature than the phase transition where the free energy curves cross. This explains also why no signatures of the phase transition is seen on the specific heat hump itself when $J_2$ crosses the phase boundary out of the K phase.

For the $J_3=0$ curve in Fig.~\ref{fig:J3zero} for $J_2 \approx 1/8$, we see a tendency of M being favored over K at finite temperatures, i.e. the M phase extends slightly into the region where K is the minimum of $J_{\qv}$. From Fig.~\ref{fig:tcslices}, we find similar behavior at finite $J_3$ for $J_3 \lesssim 0.10$.

To gain further insights on the behavior observed in Fig.~\ref{fig:tcslices}, we compute in Fig.~\ref{fig:hump_entro} the spin-wave entropy along the K-KM and K-$\Gamma$K borders using linear spin-wave theory~\cite{Seabra2016,Glittum2021}. As the energies are equal on the borders, the state with the highest spin-wave entropy should be favored close to $T=0$. For $1/15 < J_3 \lesssim 0.1$, we find that the KM state has higher entropy than the K state, while for $J_3 \gtrsim 0.1$, the K state has higher entropy than either the KM state ($0.1 \lesssim J_3 < 1/8$) or the $\Gamma$K state ($J_3 > 1/8$). For $-1 < J_3 < 1/15$, the K phase borders the M phase. The M phase, due to its collinearity, has a higher number of zero modes and thus highest entropy. We find that the low-$T$ phase in Fig.~\ref{fig:tcslices} at the K phase boundary is well explained by spin-wave entropy for all values of $J_3$.

\begin{figure}[]
\begin{center}
\includegraphics[width=0.5\textwidth]{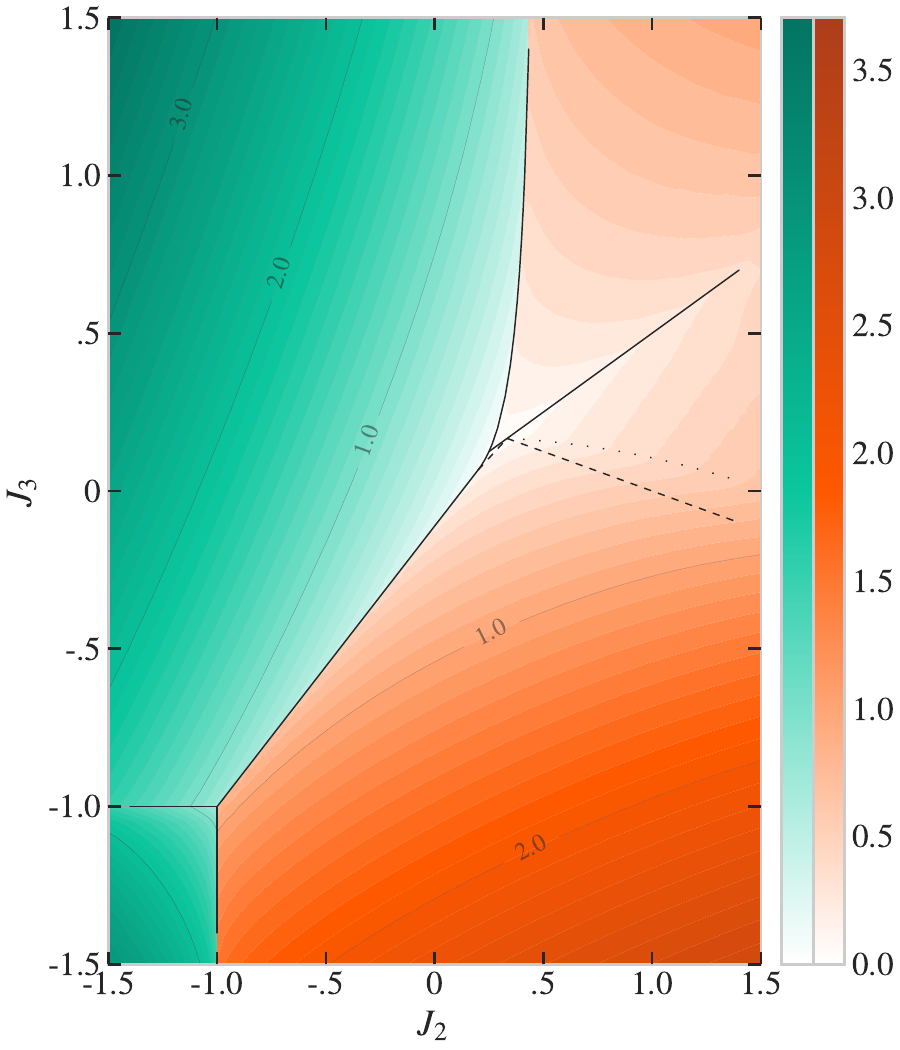}
\caption{Contour plot of $\THUMP$ (green) and $T_c$ (orange). The $T=0$ phase diagram from Fig.~\ref{fig:phasediagram_zeroT} is indicated with black solid and dashed lines. The phase transitions at $T_c$ are all discontinuous (of varying strength). The black dotted curve highlights a ``kink'' feature seen in the contours. The temperature interval between the contours is $0.1$ and contours $\THUMP,T_c = \{1,2,3\}$ are made slightly darker. $L=300$.}
\label{fig:tcs}
\end{center}
\end{figure}

Figure~\ref{fig:tcs} shows a contour plot of $\THUMP$ (green) and the highest ordering temperature $T_c$ (orange) over a large region of couplings.
All phase transitions are discontinuous, and from looking at the corresponding structure factors (not shown), we find that the low-temperature phase corresponds to single-$\qv$ states throughout the full phase diagram.
The zero-temperature phase boundaries are superimposed on the contour plot. They match well with valleys in the $\THUMP$, $T_c$ landscape, except for the continuous phase boundary between the M and $\Gamma$M phases (dashed line) which lies slightly below a curve (dotted) of contour ``kinks''. This indicates that the M phase also extends into the $\Gamma$M phase at finite temperatures so that the phase transition from the disordered state goes into a state with structure factor peaks at one of the M points. In contrast to the phase boundary between K and M there is no additional phase transition from M into $\Gamma$M at a lower temperature. Instead, the peaks in the structure factor move continuously towards one of the minima of $J_{\qv}$ along the $\Gamma$M-line as the temperature is lowered. For $J_3=0$ this crossover region is indicated by the dashed curve in Fig.~\ref{fig:J3zero}.

\begin{figure*}[]
\begin{center}
\includegraphics[width=\textwidth]{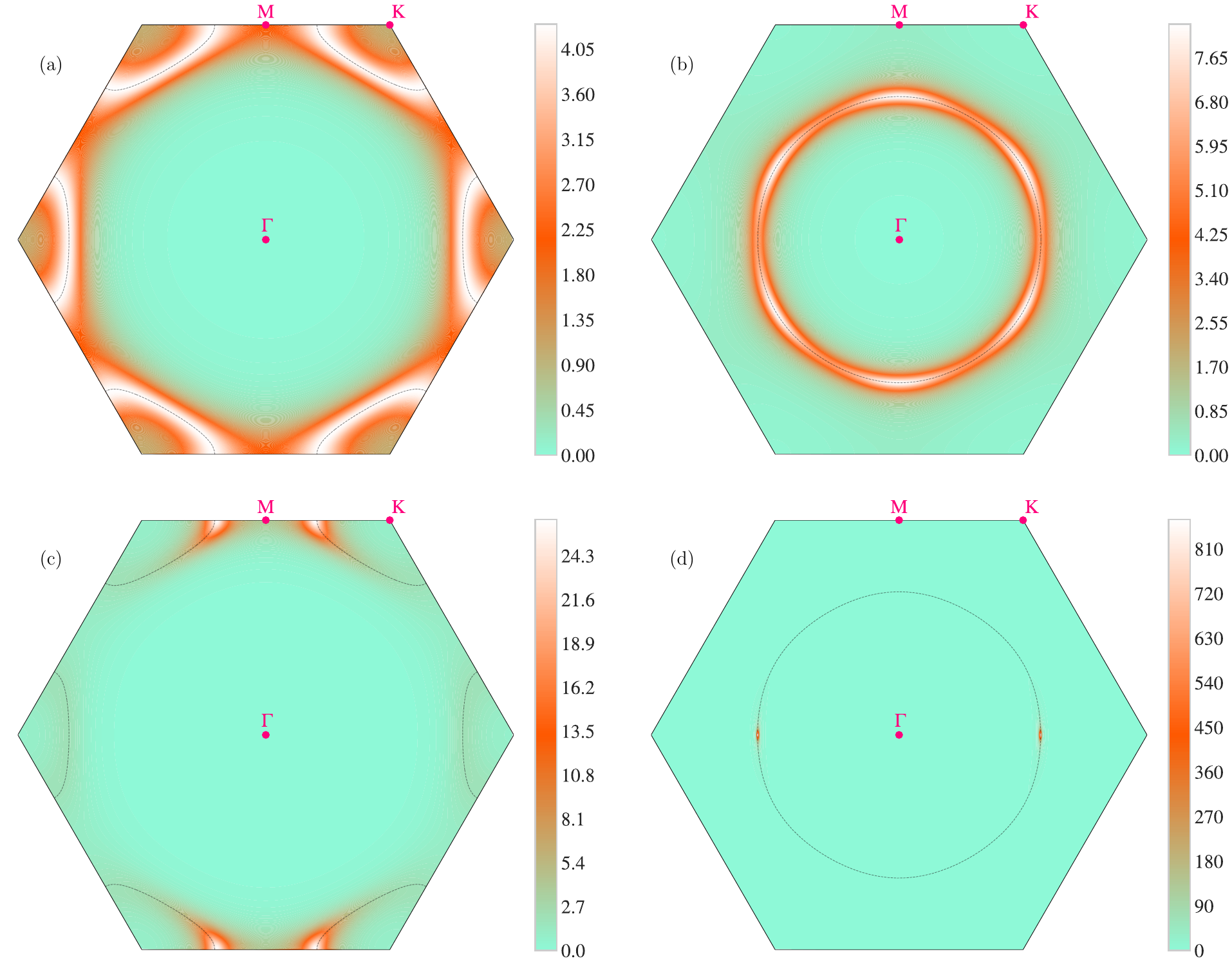}
\caption{Contour plot of the static structure factor for (a) $J_2=0.3$, $T=0.0295$, $L=300$, (b) $J_2=1$, $T=0.2270$, $L=394$, (c) $J_2=0.3$, $T=0.0277$, $L=300$, and (d) $J_2=1$, $T=0.2237$, $L=394$. The dashed curves indicate the minimum of $J_{\qv}$.}
\label{fig:modulation}
\end{center}
\end{figure*}

\subsection{Spiral spin liquids \label{sec:ring}}
Along the line $J_3 = J_2 /2$, the minima of $J_{\qv}$ form a continuous ``ring'' when $J_2 > 1/4$. This ring minimum is centered around K for $1/4 < J_2 < 1/3$ and around $\Gamma$ for $J_2 >1/3$. For $J_2 = 1/3$, it forms a hexagon with corners at the M points.

The ring minimum signifies a continuous degeneracy of spin spirals with different $\qv$ vector orientations. At low temperatures, but still above any ordering phase transitions, one expects the spin configuration to be a liquid mainly consisting of different single-$\qv$ spiral domains with momenta belonging to the ring, and the static structure factor will therefore be peaked for the ring momenta. While all single-$\qv$ spiral states on the ring are degenerate, at finite temperatures, entropy effects and interactions between different spiral domains (domain walls) will make the intensity along the ring uneven. Fig.~\ref{fig:modulation} (upper panels) show the static structure factor obtained by NBT just above the phase transition for two cases where $J_3=J_2/2$, one where the ring is around K ($J_2=0.3$) and another where the ring is around $\Gamma$ ($J_2=1$). In the latter case it is clear that the ring intensity is maximal on the $\Gamma$M lines, while the former shows a more even intensity distribution. Yet, a closer inspection reveals slightly higher intensities on the KM lines.

\begin{figure}[]
\begin{center}
\includegraphics[width=0.5\textwidth]{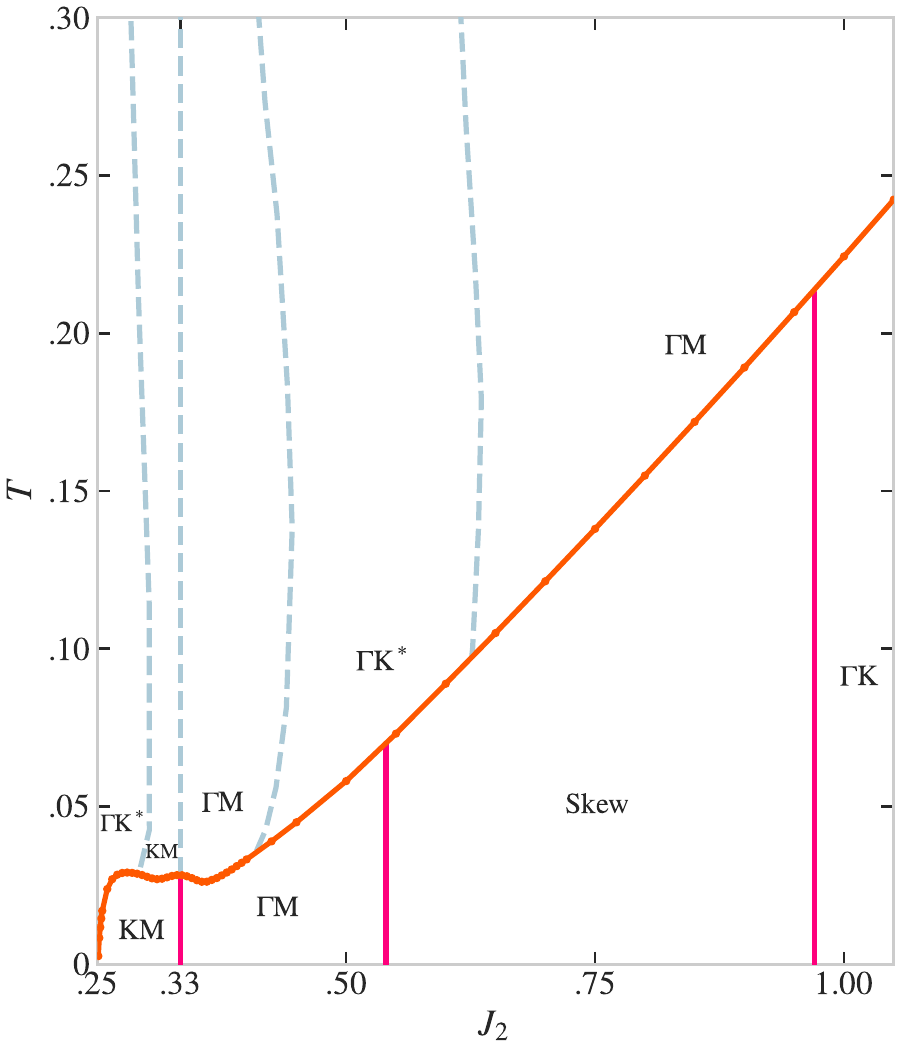}
\caption{Phase transition temperature $T_c$ (orange dots) vs. $J_2$ along the line $J_3=J_2/2$ for $J_2>1/4$. These phase transitions are all discontinuous, approaching a continuous transition as $J_2 \to 1/4$ from above. The regions separated by opalite blue dashed lines above $T_c$ indicate the direction of maximal intensity in the ring-like static structure factor. $\Gamma {\rm K}^*$ means that the maxima are located on, or close to, the $\Gamma {\rm K}$ lines. The regions below $T_c$ indicate the direction of broken symmetry. ``Skew" means a low-symmetry temperature dependent direction. $L=300$.}
\label{fig:ringmaxintensity}
\end{center}
\end{figure}

We have mapped out the directions of maximal ring intensity in the static structure factor along the line $J_3=J_2/2$ for $J_2>1/4$. The results are shown in Fig.~\ref{fig:ringmaxintensity}. The orange data points show $T_c$, and the opalite blue dashed lines above $T_c$ separate $T$-$J_2$ regions with different directions of maximal ring intensity. These regions are weakly temperature dependent, showing that the direction of maximal ring intensity may change as a function of temperature.

The phase transition at $T_c$ is discontinuous and breaks the lattice-rotational symmetry. We find that as $J_2$ approaches $1/4$, the entropy discontinuity (latent heat) becomes weaker and the phase transition approaches a continuous transition with $T_c$ going to zero.

In Fig.~\ref{fig:modulation} (lower panels), we show the static structure factor obtained by NBT just below the phase transition. It is clear that the phase transitions ruin the rotational symmetry of the spiral spin liquids. Note however that at temperatures just below $T_c$ some remnants of the ring may be visible, as seen in Fig.~\ref{fig:modulation} (c), which is similar to the broken symmetry arc seen in the FM model~\cite{Glittum2021}. Fig.~\ref{fig:modulation} (d) shows a structure factor more consistent with single-$\qv$ order (but broadened due to finite temperature). Note also that the peak in the low-temperature phase does not necessarily coincide with the maximal ring intensity above $T_c$, here illustrated by Figs.~\ref{fig:modulation} (b) and (d).

In Fig.~\ref{fig:ringmaxintensity}, we show also the directions along the ring where the static structure factor peaks in the symmetry-broken phase.  For $1/4 <J_2 < 1/3$ these maxima lie on one of the zone boundary KM directions
and for large values $J_2>0.97$ they lie on one of the $\Gamma \rm{K}$ lines, consistent with the lower panels of Fig.~\ref{fig:modulation}. For $1/3 < J_2 < 0.54$ the broken symmetry maxima lie on one of the $\Gamma \rm{M}$ directions, and for $0.54 < J_2 < 0.97$ they generally lie on a low-symmetry temperature-dependent direction marked ``Skew'' in Fig.~\ref{fig:ringmaxintensity}. This skew phase breaks also all mirror symmetries of the lattice.

\begin{figure}[]
\begin{center}
\includegraphics[width=0.5\textwidth]{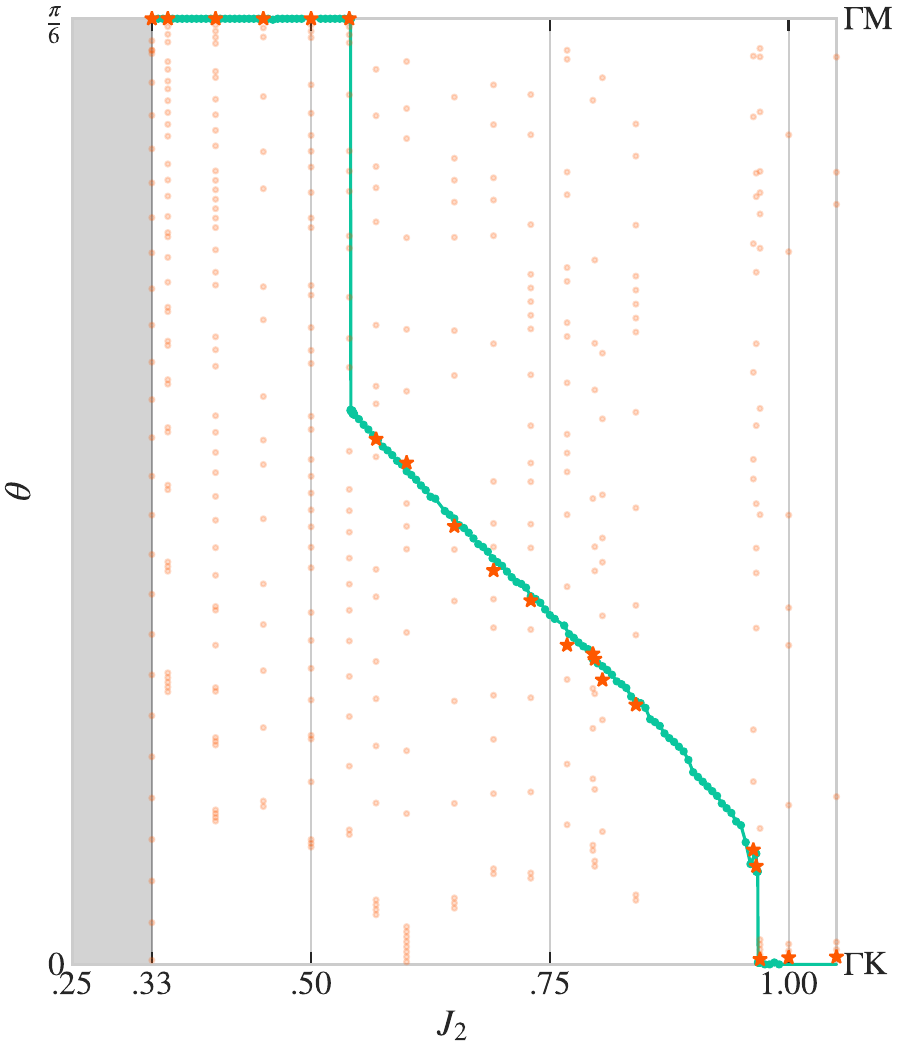}
\caption{Green: Direction $\theta$ (angle with $\Gamma$K) of maximal linear spin-wave entropy along the ring minimum for $J_2 \geq 1/3$ for $L = 1000$.  Orange: Dots show different $\qv$'s which are energetically closer than $10^{-6}$ to the ring minimum. Stars mark the states with minimal free energy from NBT. Various system sizes. Grey: For $1/4 < J_2<1/3$, both linear spin waves and NBT favor the intersection of the ring minimum and the KM line.}
\label{fig:spinwaves_ring}
\end{center}
\end{figure}

For the lowest temperatures, where the order is a single domain single-$\qv$ spiral state, the entropy is governed by spin wave fluctuations. It is this entropy that leads to the selection of a particular broken symmetry ground state at small, but finite, temperature by the order-by-disorder scenario~\cite{Villain1980,Henley1989,Chandra1990}.
We have calculated the spin-wave entropy associated with the different ordering $\qv$-vectors along the ring minimum using linear spin-wave theory~\cite{Seabra2016,Glittum2021}. The wave vectors of maximal entropy are shown in Fig.~\ref{fig:spinwaves_ring} (green curve). They are parametrized by their direction $\theta$, the angle with the $\Gamma$K-direction, so that $\theta = \pi/6$ is along $\Gamma$M.
In the region $1/4 < J_2 < 1/3$ the ring is centered on the K points and the entropy is maximal along the Brillouin zone boundary (KM lines), going from K at $J_2=1/4$ to M at $J_2=1/3$. For higher values of $J_2$, the ring is centered around $\Gamma$. We then find the entropy to be maximal in the $\Gamma$M direction for $1/3 < J_2 \lesssim 0.54$ and the $\Gamma$K direction for $J_2 \gtrsim 0.97$.  In the region $0.54 < J_2 < 0.97$, $\theta$ decreases continuously from $0.31$ at $J_2 \approx 0.54$ to $0.051$ at $J_2 \approx 0.97$. The changes in $\theta$ from $\pi/6$ ($\Gamma$M) to $0.31$ at $J_2 \approx 0.54$ and from $0.051$ to $0$ ($\Gamma$K) at $J_2 \approx 0.97$ are on the other hand both discontinuous.

The comparison of these entropy maxima with the ordering wave vectors obtained in NBT is delicate for a lattice of finite size. This is because the set of $\qv$-points that lie energetically close to the ring minimum varies greatly as the ring changes radius when $J_2$ is altered. The light orange dots in Fig.~\ref{fig:spinwaves_ring} show the directions of the $\qv$-points that deviate no more than $10^{-6}$ from the minimum energy for selected $J_2$ values. In our selection of $J_2$ values and system sizes for NBT we have made sure that there is at least one $\qv$-point which is close to the orientation of the maximal entropy (green curve). The orange stars in Fig.~\ref{fig:spinwaves_ring} show the direction of the ordering wave vector giving smallest free energy in NBT at very low temperatures ($\sim 10^{-4}$). They are consistent with the directions of maximal spin-wave entropy.

Note that for the region $0.85 < J_2 < 0.95$, we show no data points from NBT. In this region the spin-wave entropy difference between $\Gamma$K and the favored direction (green curve) is small ($\sim 10^{-3}$). With such small entropy differences, the finite size effect which leads to small energy differences ($\sim 10^{-6}$) of the $\qv$-points on the ring becomes increasingly important and one must consider higher temperatures ($\sim 10^{-3}$) in order for entropy, as opposed to the energy, to determine which spiral state is stabilized. At such ``high'' temperatures, the entropy is likely not determined purely by linear spin-wave fluctuations. Thus, we do not have any reliable data in this region.

\section{Discussion \label{sec:discussion}}

As seen in Fig.~\ref{fig:humpJ2}, NBT gives a specific heat for the pure nearest-neighbor model which has a broad asymmetric hump at $\THUMP \approx 0.367$. While the existence of the broad hump is in accordance with Monte Carlo simulations~\cite{Kawamura2010}, NBT overestimates its peak temperature by 16\%.
This is in line with NBT typically overestimating critical temperatures by 10-20\%~\cite{Syljuasen2019}.

Kawamura and Myashita's Z$_2$ vortex theory~\cite{Kawamura1984} interprets features of the specific heat hump as a signature of a finite-temperature topological transition driven by unbinding of Z$_2$ vortices~\cite{Kawamura2010, Aoyama2020}. This scenario predicts two distinct length scales and a weak essential singularity in the specific heat. Such unbinding of Z$_2$ vortices is an effect relying on the topological nature of planar spin configurations with three sublattices. As NBT is based on a large-$N_s$ expansion, where $N_s$ is the number of spin components, it is unlikely that it captures correctly topological effects that crucially depend on the precise value of $N_s$. Nevertheless, NBT gives a clear hump feature in the specific heat at approximately the correct temperature, and we cannot rule out that NBT may capture local energy consequences as defects unbind.
We show in Fig.~\ref{fig:corrlength} that the temperature of this hump is associated with the onset of spin correlations with a correlation length growing exponentially below $\THUMP$; $\xi \sim e^{B/T}$. The constant $B$ for the nearest neighbor antiferromagnet has previously been calculated from an effective SO(3) nonlinear sigma model~\cite{Chakravarty1988,Dombre1989,Azaria1992}, and is proportional to the spin stiffness $\rho_s$ (for twists about axes lying in the plane of the spins); $B = c \rho_s$ where $c=2\pi(1+\pi/2) \approx 16.15$. Generalizing the spin stiffness result in Ref.~\onlinecite{Dombre1989} to further neighbor couplings, we find $\rho_s = \f{\sqrt{3}}{4} \left( J_1 - 6 J_2 + 4J_3 \right)$. The exchange coupling dependence of $B$ observed in Fig.~\ref{fig:corrlength} is consistent with that of the spin stiffness, but we get a considerably lower value of the proportionality constant $c \approx 11$, a factor $1.5$ down from the result in Ref.~\onlinecite{Azaria1992}. This discrepancy is not a consequence of NBT overestimating temperatures, as that would result in an overestimation of $c$. We note that previous Monte Carlo studies concluded that the finite size behavior of the spin stiffness of the triangular Heisenberg antiferromagnet is very well described by the nonlinear sigma model, but were not able to verify the actual value of $B$ because of insufficient system sizes~\cite{Southern1993,Wintel1995}.

The exponential growth lends support to the notion that the specific heat hump indicates a would-be phase transition if it was not forbidden by the Mermin-Wagner theorem~\cite{Mermin1966}. We note that a broad specific heat hump also exists for the Heisenberg model on the square lattice~\cite{Colot1983}.

Despite the Mermin-Wagner theorem, there can still be finite-temperature phase transitions that break {\em discrete} lattice symmetries. We have mapped out critical temperatures for these phase transitions, which are all discontinuous (except for the multicritical point), as well as the peak temperature of the specific heat humps for the full range of $J_2$-$J_3$ couplings. This can aid the inference of specific coupling values from experimental measurements of either the magnetic specific heat or phase transition temperatures.

Our results are in general consistent with the Monte Carlo results in Ref.~\onlinecite{Mohylna2022}, which mainly focused on investigating the low-temperature order of a single $(J_2,J_3)$ point within each phase of the zero-temperature phase diagram. Like us, they find that the low-temperature phase corresponds to a single-$\qv$ state in all phases. This includes also the M phase, which is expected from entropy arguments as the single-$\qv$ stripe state is collinear, while the multi-$\qv$ tetrahedral state is non-coplanar~\cite{Messio2011}.

The finite-temperature phase diagram of the model studied here contains a multicritical point at $J_2=1/4$, $J_3=1/8$ and $T_c=0$ where four phases meet: the $T=0$ uniformly ordered AF 120{\degree} state (K), two incommensurate spiral phases ($\Gamma$K and KM), and the disordered phase.
Our results show that going away from the multicritial point along the line $J_3=J_2/2$ as $J_2>1/4$, both the critical temperature and latent heat increase continuously from zero. The phase transitions along this ring-minimum line separate disordered spiral spin liquid states from low-temperature ordered states that break the lattice-rotational symmetry. The pattern of symmetry breaking follows the order-by-disorder scenario and agrees with that predicted by calculations of the entropy within the linear spin-wave approximation.

These results partially agree with Ref.~\onlinecite{Mohylna2022}. They investigate two points along the ring minimum line: $J_2 = 0.3$ and $J_2 = 0.5$. For $J_2 = 0.3$, our results agree with theirs, with the low-temperature order being along KM. Their critical temperature is $T_c \approx 0.02$, while ours is $T_c = 0.028$. For $J_2 = 0.5$, Ref.~\onlinecite{Mohylna2022} finds the low-temperature order to be a skew state close to $\Gamma$K. This is not consistent with our results giving $\Gamma$M, which is supported by spin-wave entropy calculations. Also their critical temperature is $T_c \approx 0.043$, while ours is $T_c = 0.058$. We believe the discrepancy in the low-temperature phase at $J_2 = 0.5$ between our results and those in Ref.~\onlinecite{Mohylna2022} is due to their limited system sizes. The available $\qv$'s lying on the ring minimum is strongly dependent on system size, and their system sizes do not access the $\qv$ lying on $\Gamma$M (within energy differences of $10^{-6}$). This might also explain why the overestimation of the critical temperatures is higher than usual.

For a FM nearest neighbor interaction, a spiral spin liquid was observed recently in the $S=3/2$ quasi-two-dimensional delafossite-like AgCrSe$_2$~\cite{Andriushin2025}, where the maximal neutron scattering intensites are oriented along the $\Gamma$K lines. There, the exchange couplings deviate from the line $J_3=J_2/2$, with $J_3$ being slightly higher. This cause the energy minima to be on the $\Gamma$K lines, which increase the intensities of the static structure factor in the $\Gamma$K-directions. However, as we have shown in Fig.~\ref{fig:ringmaxintensity} (for AF couplings), entropy and interactions of spirals can also select a particular modulation of the structure factor intensity, even on the line $J_3=J_2/2$, where all spiral $\qv$'s have the same energy. By continuity, we expect that this entropic effect will extend these regions outside of the $J_3=J_2/2$ line, until $\Gamma$K or $\Gamma$M/KM spiral-$\qv$s become so energetically favored that they will dominate below a certain temperature. It is therefore actually possible to have a spiral spin liquid modulated along $\Gamma$K($\Gamma$M/KM) despite the minimal energy of $J_{\qv}$ being along $\Gamma$M/KM($\Gamma$K).

In addition to the triangular lattice Heisenberg ferromagnet, spiral spin liquids were previously predicted theoretically for the diamond~\cite{Bergman2007} and honeycomb lattice~\cite{Okumura2010} Heisenberg antiferromagnets, and the square lattice XY ferromagnet~\cite{Yan2022, Gonzalez2024}, also there with further-neighbor interactions. Here, we have shown that also the archetypical triangular lattice antiferromagnet hosts spiral spin liquids when including second- and third-nearest neighbor couplings. On our wish list is to investigate the real space configurations of this Heisenberg spiral spin liquid, to better understand the underlying excitations, which in the XY case are known to be momentum vortices~\cite{Yan2022}.

Our model is the classical analog of the suggested models for $S=1/2$ delafossite compounds~\cite{Scheie2024,Scheie2024_2, Xie2023}, where further-neighbor couplings ($J_2$) are often sizeable. Our work is also relevant for triangular Heisenberg magnets of higher spin. Such ``classical'' examples include layered compounds such as the $S=5/2$ materials RbFe(MoO$_4$)$_2$~\cite{Svistov2003}, Rb$_4$Mn(MoO$_4$)$_3$~\cite{Ishii2011}, NH$_4$Fe(PO$_3$F)$_2$~\cite{Mohanty2025}, and h-Lu$_{0.47}$Sc$_{0.53}$FeO$_3$~\cite{Yano2024}, the $S=2$ material h-Lu$_{0.3}$Y$_{0.7}$MnO$_3$~\cite{Yano2024}, and the $S=3/2$ materials LiCrO$_2$~\cite{Kadowaki1995}, and LiCrS$_2$~\cite{Ushakov2013}.

The three $S=5/2$ materials Ba$_3$MnNb$_2$O$_9$, Ba$_8$MnNb$_6$O$_{24}$ and Ba$_3$MnSb$_2$O$_9$ are ``classical'' triangular antiferromagnets which have been observed to order in the $120\degree$ phase at low temperatures\markup{, exhibiting varying degrees of out‑of‑plane canting}~\cite{Lee2014,Jiao2022,Rawl2019,Tian2014,Shu2023}. Despite their similar structures, these materials have quite different Curie-Weiss temperatures, meaning that the exchange couplings of the materials are likely to be different. We note that the combination of in-plane $J_1$, $J_2$ and $J_3$ must be chosen to lie within the K region of the ground state phase diagram in Fig.~\ref{fig:phasediagram_zeroT} to be consistent with the $120\degree$ phase for the lowest temperatures.

The $S=2$ material FeGa$_2$S$_4$ is a triangular layered antiferromagnet which shows short-range in-plane magnetic correlations in a $\Gamma$K state at low temperatures with $q^x_{\Gamma {\rm K}} = 2.79$. The in-plane couplings are suggested to be described by a $J_1$-$J_2$-$J_3$ classical Heisenberg model with $J_1 = 1.7$ meV, $J_2 = 0.9$ meV and $J_3=0.8$ meV~\cite{Guratinder2021}, although there are considerable substitutional disorder of Fe and Ga atoms~\cite{Guratinder2021}. We find that the suggested couplings lie within the $\Gamma$K phase (Fig.~\ref{fig:phasediagram_zeroT}) and correspond to $q^x_{\Gamma {\rm K}} = 2.62$.

Lastly, we note that two proposed QSL regimes of the $S=1/2$ \textit{quantum} model both coincide with classical phase boundaries~\cite{Gong2019}: The QSL proposed for the $J_1$-$J_2$ model is suggested to be a broadening of the classical K-M boundary, also when extended to the $J_1$-$J_2$-$J_3$ model, and a chiral QSL occurs near the $J_3=J_2/2$ line for $0.2 < J_2 < 0.5$.
These two QSL regimes are thus realised where the classical ordering temperature is suppressed, as seen in Fig.~\ref{fig:tcs}. We also note that these regimes meet near the multicritical point of the classical model. This illustrates the importance of studying thermal behavior of classical models, not only to understand thermal properties, but also to provide guidance on where quantum fluctuations are likely to be important.

\section{Conclusion \label{sec:conclusion}}

Seeing the wood for the trees, the NBT provides a quantitative description of the finite-temperature behavior of the classical triangular Heisenberg antiferromagnet with extended couplings.
In addition to the conventional 120{\degree} phase, the second- and third-nearest neighbor interactions give rise to phases that break discrete rotational lattice symmetries and also finite temperature spiral spin liquids. It is our hope that the quantitative finite temperature results presented here can aid the deduction of possible values of exchange couplings from experimental measurements of magnetic specific heat, phase transition temperatures and neutron scattering structure factors.

\section{Acknowledgements}
C.G. thanks T.A. Swift for setting the stage for the manuscript's style. The computations were performed on resources provided by Sigma2 - the National Infrastructure for High Performance Computing and Data Storage in Norway, and on the Fox supercomputer at the University of Oslo.

\vspace{3mm}
The data that support the findings of this study are openly available at the following URL/DOI: \url{10.5281/zenodo.18 667 960}.

\bibliographystyle{iopart-num}
\bibliography{hump}

\end{document}